# Congestion Avoidance in Computer Networks With a Connectionless Network Layer

Raj Jain, K. K. Ramakrishnan, Dah-Ming Chiu
Digital Equipment Corporation
550 King St. (LKG1-2/A19)
Littleton, MA 01460

**Abstract**

Widespread use of computer networks and the use of varied technology for the interconnection of computers has made congestion a significant problem.

In this report, we summarize our research on congestion avoidance. We compare the concept of *congestion avoidance* with that of *congestion control*. Briefly, congestion control is a recovery mechanism, while congestion avoidance is a prevention mechanism. A congestion control scheme helps the network to recover from the congestion state while a congestion avoidance scheme allows a network to operate in the region of low delay and high throughput with minimal queuing, thereby preventing it from entering the congested state in which packets are lost due to buffer shortage.

A number of possible alternatives for congestion avoidance were identified. From these alternatives we selected one called the *binary feedback scheme* in which the network uses a single bit in the network layer header to feed back the congestion information to its users, which then increase or decrease their load to make optimal use of the resources. The concept of *global optimality* in a distributed system is defined in terms of efficiency and fairness such that they can be independently quantified and apply to any number of resources and users.

The proposed scheme has been simulated and shown to be globally efficient, fair, responsive, convergent, robust, distributed, and configuration-independent.

## 1 INTRODUCTION

Congestion in computer networks is becoming a significant problem due to increasing use of the networks, as well as due to increasing mismatch in link speeds caused by intermixing of old and new technology. Recent technological advances such as local area networks (LANs) and fiber optic LANs have resulted in a significant increase in the bandwidths of computer network links. However, these new technologies must coexist with the old low bandwidth media such as the twisted pair. This heterogeneity has resulted in mismatch of arrival and service rates in the intermediate nodes in the network, causing increased queuing and congestion.

We are concerned here with *congestion avoidance* rather than *congestion control*. Briefly, a congestion avoidance scheme allows a network to operate in the region of low delay and high throughput. These schemes prevent a network from entering the congested state in which the packets are lost. We will elaborate on this point in the next section where the terms flow control, congestion control, and congestion avoidance will be defined and their relationship to each other discussed.

We studied a number of alternative schemes for congestion avoidance. Based on a number of requirements described later in this report, we selected an alternative called the *binary feedback scheme* for detailed study. This scheme uses only a single bit in the network layer header to feed back the congestion information from the network to users, which then increase or decrease their load on the network to make efficient and fair use of the resources. We present precise definitions of efficiency and fairness that can be used for other distributed systems as well.

This report is a summary of our work in the area of congestion avoidance in connectionless networks. We have tried to make this summary as self-contained and brief as possible. For further information, the reader is encouraged to read detailed reports in [16, 22, 4, 23].

# 2 CONCEPTS

In this section we define the basic concepts of flow control, congestion control, and congestion avoidance.

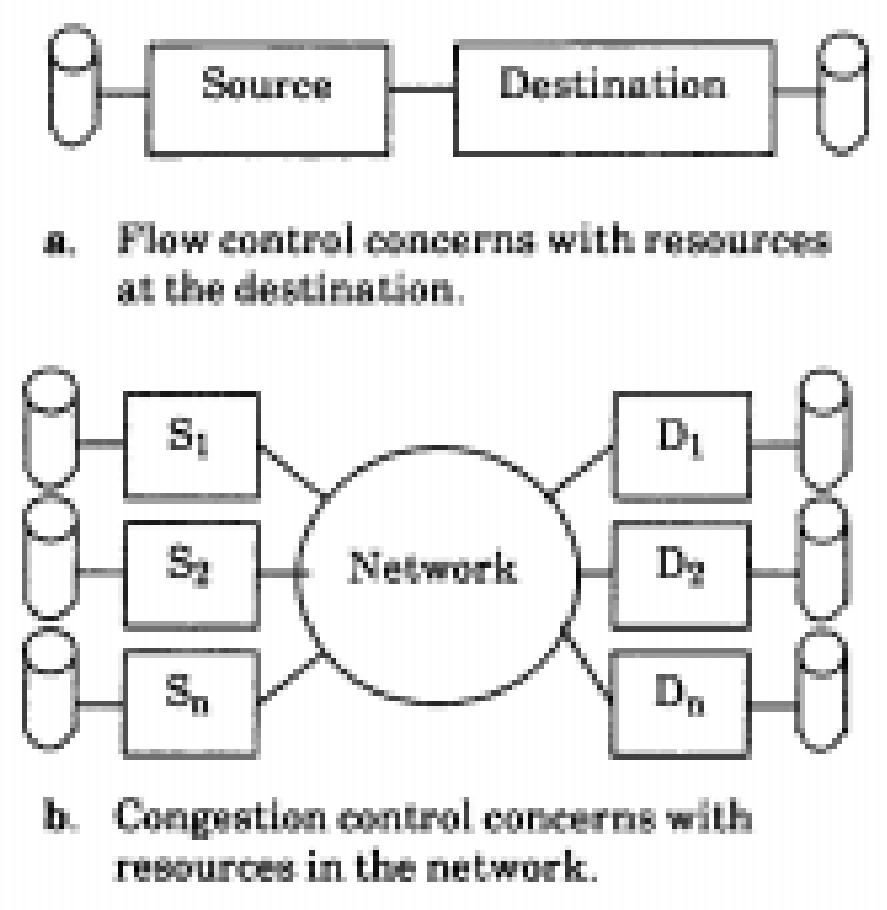

Figure 1:

Consider the simple configuration shown in Figure 1a, in which two nodes are directly connected via a link. Without any control, the source may send packets at a rate too fast for the destination. This may cause buffer overflow at the destination, leading to packet losses, retransmissions, and degraded performance. A **flow control** scheme protects the destination from being flooded by the source.

Some of the alternatives that have been described in the literature are window flow-control, Xon/Xoff [7], rate flow-control [5], etc. In the window flow-control scheme, the destination specifies a limit on the number of packets that the source may send without further permission from the destination.

Let us now extend the configuration to include a communication subnet (see Figure 1b) consisting of routers and links that have limited memory, bandwidth, and processing speeds. Now the source must not only obey the directives from the destination, but also from all the routers and links in the network. Without this additional control the source may send packets at a rate too fast for the network, leading to queuing, buffer overflow, packet losses, retransmissions, and performance degradation. A **congestion control** scheme protects the network from being flooded by its users (transport entities at source and destination nodes).

In connection-oriented networks the congestion problem is generally solved by reserving the resources at all routers during connection setup. In connectionless networks it can be done by explicit messages (choke packets) from the network to the sources [19], or by implicit means such as timeout on a packet loss. In [15, 13, 21], a number of alternatives have been discussed and a timeout-based scheme has been analyzed in detail.

Traditional congestion control schemes help improve the performance after congestion has occurred. Figure 2 shows general patterns of response time and throughput of a network as the network load increases. If the load is small, throughput generally keeps up with the load. As the load increases, throughput increases. After the load reaches the network capacity, throughput stops increasing. If the load is increased any further, the queues start building, potentially resulting in packets being dropped. Throughput may suddenly drop when the load increases beyond this point and the network is said to be *congested*. The response-time curve follows a similar pattern. At first the response time increases little with load. When the queues start building up, the response time increases linearly until finally, as the queues start overflowing, the response time increases drastically.

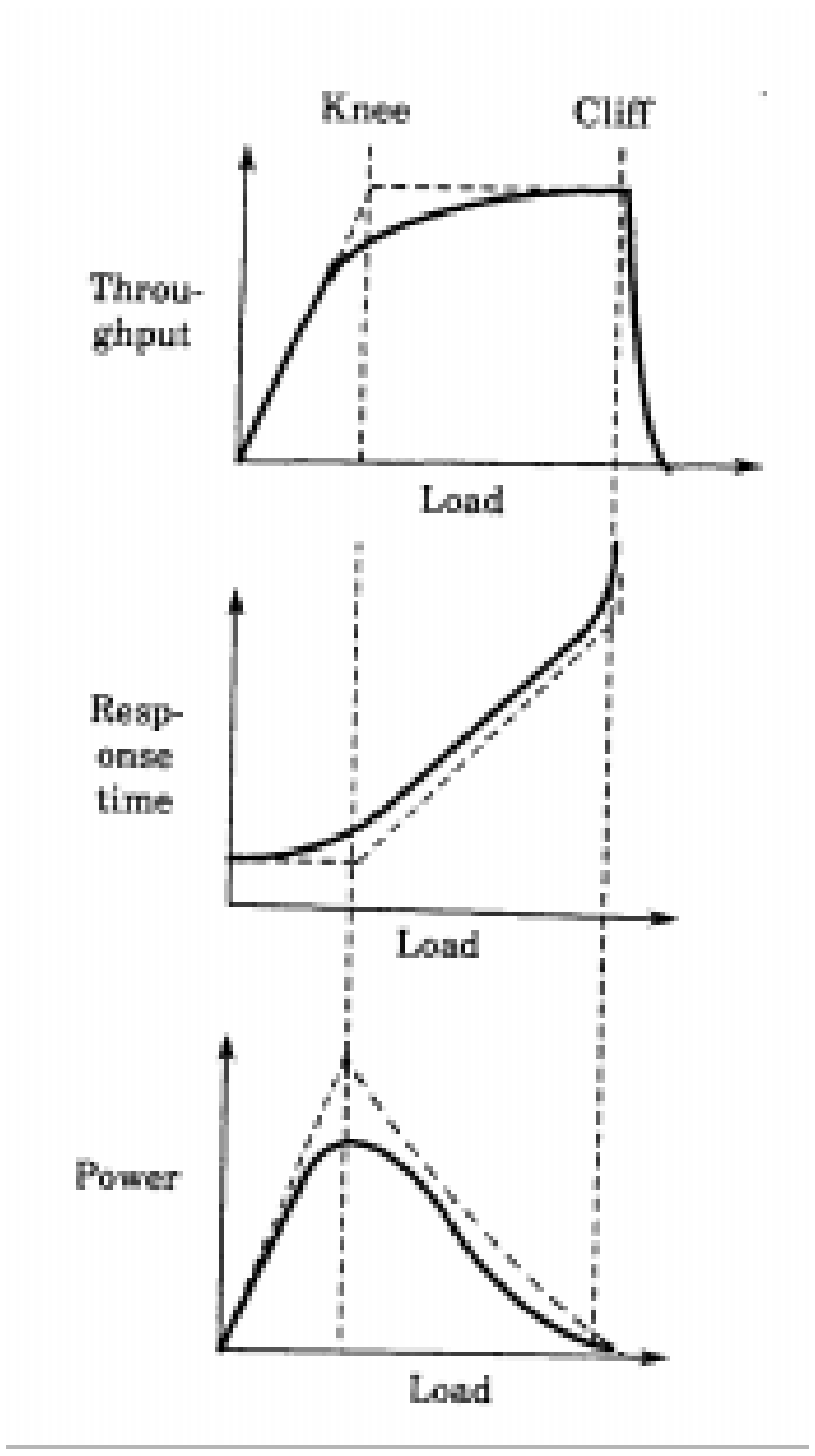

Figure 2:

The point at which throughput approaches zero is called the point of *congestion collapse*. This is also the point at which the response time approaches infinity. The purpose of a congestion control scheme (such as [15, 3]) is to detect the fact that the network has reached the point of congestion collapse resulting in packet losses, and to reduce the load so that the network returns to an uncongested state.

We call the point of congestion collapse a **cliff** due the fact that the throughput falls off rapidly after this point. We use the term **knee** to describe the point after which the increase in the throughput is small, but after which a significant increase in the response time results.

A scheme that allows the network to operate at the knee is called a **congestion avoidance** scheme, as distinguished from a *congestion control* scheme that tries to keep the network operating in the zone to the left of the cliff. A properly designed congestion avoidance scheme will ensure that the users are encouraged to increase their traffic load as long as this does not significantly affect the response time and are required to decrease them if that happens. Thus, the network load oscillates around the knee. Congestion control schemes are still required, however, to protect the network should it reach the cliff due to transient changes in the network.

The distinction between congestion control and congestion avoidance is similar to that between deadlock recovery and deadlock avoidance. Congestion control procedures are curative and the avoidance procedures are preventive in nature. The point at which a congestion control scheme is called upon depends upon the amount of memory available in the routers, whereas the point at which a congestion avoidance scheme is invoked is independent of the memory size.

We elaborate further on these concepts in [16].

# 3 ALTERNATIVES

Congestion control and congestion avoidance are dynamic system control issues. Like all other control schemes they consist of two parts: a feedback mechanism and a control mechanism. The feedback mechanism allows the system (network) to inform its users (sources or destinations) of the current state of the system, and the control mechanism allows the users to adjust their loads on the system.

The problem of congestion control has been discussed extensively in the literature. A number of feedback mechanisms have been proposed. If we extend those mechanisms to operate the network around the knee rather than the cliff, we obtain congestion avoidance mechanisms. For the feedback mechanisms we have the following alternatives:

1. Congestion feedback via packets sent from routers to sources

2. Feedback included in the routing messages exchanged among routers
3. End-to-end probe packets sent by sources
4. Each packet containing a congestion feedback field filled in by routers in packets going in the reverse direction– *reverse feedback*
5. Each packet containing a congestion feedback field filled in by routers in packets going in the forward direction– *forward feedback*

The first alternative is popularly known as *choke packet* [19] or *source quench message* in ARPAnet [20]. It requires introducing additional traffic in the network during congestion, which may not be desirable.

The second alternative, increasing the cost (used in updating the forwarding database) of congested paths, has been tried before in ARPAnet's delay-sensitive routing. The delays were found to vary too quickly, resulting in a high overhead [18].

The third alternative, probe packets, also suffers from the disadvantage of added overhead, unless probe packets have a dual role of carrying other information in them. If the latter were the case, there would be no reason not to use every packet going through the network as a probe packet. We may achieve this by reserving a field in the packet that is used by the network to signal congestion. This leads us to the last two alternatives.

The fourth alternative, reverse feedback, requires routers to piggyback the signal on the packets going in the direction opposite the congestion. This alternative has the advantage that the feedback reaches the source faster. However, the forward and reverse traffic are not always related. The destinations of the reverse traffic may not be the cause of or even the participants in the congestion on the forward path. Also, many networks (including Digital Network Architecture, or DNA) have path-splitting such that the path from A to B is not necessarily the same as that from B to A.

The fifth alternative, forward feedback, sends the signal in the packets going in the forward direction (direction of congestion). In the case of congestion the destination either asks the source to reduce the load or returns the signal back to the source in the packets (or acknowledgments) going in the reverse direction. This is the alternative that we study here and in [22, 23].

The key architectural assumption about the networks in this study is that they use connectionless network service and transport level connections. By this we mean that a router is not aware of the transport connections passing through it, and the transport entities are not aware of the path used by their packets. There is no prior reservation of resources at routers before an entity sets up a connection. The routers cannot compute the resource demands except by observing the traffic flowing through them. Examples of network architectures with connectionless network layers are DoD TCP/IP, DNA, and ISO connectionless network service used with ISO transport class 4 [9].

# 4 PERFORMANCE METRICS

A congestion avoidance scheme is basically a resource allocation mechanism in which the subnet (set of intermediate nodes or routers) is a set of $m$ resources that has to be allocated to $n$ users (source-destination pairs). There are two parties involved in any resource allocation mechanism: the resource manager and the user. The resource manager's goal is to use the resource as efficiently as possible. Users, on the other hand, are more interested in getting a fair share of the resource. We therefore need to define efficiency and fairness.

For our current problem of congestion avoidance, the routers are our resources and therefore we use the terms *routers* and *resources* interchangeably. The concepts introduced here, however, are general and apply to other distributed resource allocation problems as well. Similarly, for the current problem, the demands and allocations are measured by packets/second (throughput), but the concepts apply to other ways of quantifying demands and allocations.

*Readers not interested in definitions of these met-*

*rics may skip to the next section on the proposed scheme.*

### 4.1 Single Resource, Single User

Consider first only one user and one resource. In this case fairness is not an issue. If the user is allowed to increase its demand (window), the throughput increases. However, the response time (total waiting time at the resource) also increases. Although we want to achieve as high a throughput as possible, we also want to keep the response time as small as possible. One way to achieve a tradeoff between these conflicting requirements is to maximize resource power [8, 17], which is defined by:

$$\text{Resource Power} = \frac{\text{Resource Throughput}^{\alpha}}{\text{Resource Response Time}}$$

Here, $\alpha$ is a constant. Generally, $\alpha = 1$. Other values of $\alpha$ can be used to give higher preference to throughput ($\alpha > 1$) or response time ($\alpha < 1$). The concepts presented in this report apply to any value of $\alpha$. However, unless otherwise specified we will assume throughout this report that $\alpha = 1$. The resource power is maximum at the knee.

For any given inter-arrival and service time distributions, we can compute the throughput at the knee. We call this the **knee-capacity** of the resource.

The **maximally efficient operating point** for the resource is its knee. The efficiency of resource usage is therefore quantified by:

$$\text{Resource Efficiency} = \frac{\text{Resource Power}}{\text{Resource Power at Knee}}$$

The resource is used at 100% efficiency at the knee. As we move away from the knee, the resource is being used inefficiently, that is, either underutilized (throughput lower than the knee-capacity) or overutilized (high response time).

### 4.2 Single Resource, Multiple Users with Equal Demands

With multiple users we have an additional requirement of fairness. The allocation is efficient as long as the total throughput is equal to the knee-capacity of the resource. However, a maximally efficient allocation may not be fair, as some users may get better treatment than others. The fairness of an allocation is a function of the amounts demanded as well as the amounts allocated. To simplify the problem, let us first consider the case of equal demands in which all users have identical demands (D). The **maximally fair allocation** then consists of equal allocations to all users, i.e., $a_i = A$ for all $i$. The fairness of any other (non-equal) allocation is measured by the following **fairness function** [11]:

$$\text{Fairness} = \frac{(\sum_{i=1}^{n} x_i)^2}{n \sum_{i=1}^{n} x_i^2} \tag{1}$$

where $x_i = a_i/D$.

This function has the property that its value always lies between 0 and 1 and that 1 (or 100%) represents a maximally fair allocation.

Notice that we use *user throughput* to measure allocations and demands because of its additivity property: total throughput of $n$ users at a single resource is the sum of their individual throughputs.

### 4.3 Single Resource, Multiple Users with Unequal Demands

Given a resource with knee-capacity of $T_{knee}$, each of the $n$ users deserves a fair share of $T_{knee}/n$. However, there is no point in allocating $T_{knee}/n$ to a user who is demanding less than $T_{knee}/n$. It would be better to give the excess to another user who needs more. This argument leads us to extend the concept of *maximally fair allocation* such that the fair share $t_f$ is computed subject to the following two constraints:

1. The resource is fully allocated:

$$\sum_{i=1}^{n} a_i = T_{knee}$$

2. No one gets more than the fair share or its demands

$$a_i = \min\{d_i, t_f\}$$

Given the knee capacity of a resource and individual user demands, the above two constraints allows us to determine the maximally fair allocation $\{A_1^*, A_2^*, \ldots, A_n^*\}$. If actual allocation $\{a_1, \ldots, a_n\}$ is different from this, we need a distance function to quantify the fairness. We do this by using the *fairness function* of equation 1 with $x_i = a_i/A_i^*$.

The efficiency of the resource usage can be computed as before by computing resource power from the resource throughput (which is given as the sum of user throughputs in this case) and the resource response time. The allocation that is 100% efficient and 100% fair is the **optimal allocation**.

We must point out that the above discussion for a single resource case also applies if there are multiple ($m$) routers but all routers are shared by all $n$ users. In this case the set of $m$ routers can be combined and considered as one resource.

### 4.4 Multiple Resources, One User

We have extended the concepts of fairness and efficiency to a distributed system with multiple resources. Let us first consider a case of a single user so that fairness is not an issue. For the subnet congestion problem, the user has a **path** $P$ passing through $m$ resources (routers) $\{r_1, r_2, \ldots, r_m\}$. The resource with the lowest service rate determines the user's throughput and is called the **bottleneck** resource. The bottleneck resource has the highest utilization (ratio of throughput to service rate) and contributes the most to user's response time. The maximally efficient operating point for the system is defined as the same as that for the bottleneck router. Thus, given a system of $m$ resources, we determine the bottleneck and define its efficiency as the **global efficiency** and its knee as the maximally efficient operating point for the system.

Global Efficiency = Efficiency of the Bottleneck Resource

Note that the global efficiency, as defined here, depends upon the response time at the bottleneck resource and not on the user response time, which is a sum of response time at $m$ resources.

### 4.5 Multiple Resources, Multiple Users

In this case, there are $n$ users and $m$ resources. The $i^{th}$ user has a path $p_i$ consisting of a subset of resources $\{r_{i1}, r_{i2}, \ldots, r_{im_i}\}$. Similarly, $j^{th}$ resource serves $n_j$ users $\{U_{j1}, U_{j2}, \ldots, U_{jn_j}\}$. The global efficiency is still defined by the bottleneck resource which is identified by the resource with the highest utilization. The problem of finding the maximally efficient and maximally fair allocation is now a constrained optimization problem as it has to take differing user paths into account. We have developed an algorithm [23] which gives the globally optimal (fair and efficient) allocation for any given set of resources, users, and paths.

Once globally optimal allocation $\{A_1^*, A_2^*, \ldots, A_n^*\}$ has been determined, it is easy to quantify fairness of any other allocation $\{a_1, a_2, \ldots, a_n\}$ by using the same fairness function as in the single resource case (equation 1) with $x_i = a_i/A_i^*$.

This fairness is called **global fairness** and the efficiency of the bottleneck resources is called the global efficiency. An allocation which is 100% globally efficient and 100% globally fair is said to be **globally optimal**. It should be pointed out that by associating efficiency with resource power (rather than user power), we have been able to avoid the problems encountered by other researchers [2, 10] in using the power metric.

Notice that we have a multi-criteria optimization problem since we are trying to maximize efficiency as well as fairness. One way to solve such problems is to combine the multiple criteria into one, for instance by taking a weighted sum or by taking a product. We chose instead to put a strict priority on the two criteria. Efficiency has a higher priority than fairness. Given two alternatives, we prefer the more efficient alternative. Given two alternatives with equal efficiency, we choose the fairer alternative.

## 5 THE PROPOSED SCHEME

We have designed a scheme that allows a network to operate at its knee. As shown in Figure 3, the scheme uses one bit called the **congestion avoidance bit** in the network layer header of the packet for feedback from the subnet to the users. A source clears the congestion avoidance bit as the packet enters the subnet. All routers in the subnet monitor their load and if they detect that they are operating above the knee, they set the congestion avoidance bit in the packets belonging to users causing overload. Routers operating below the knee pass the bit as received. When the packet is received at the destination the network layer passes the bit to the destination transport, which takes action based on the bits.

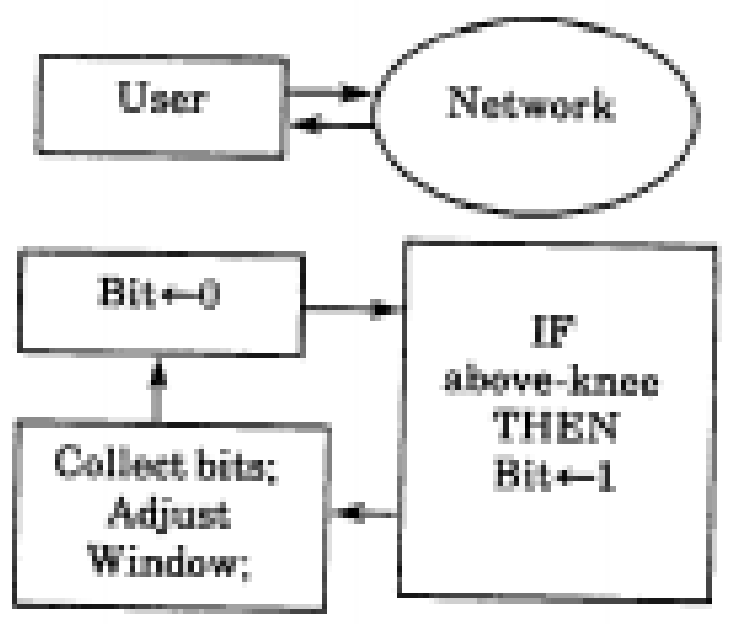

Figure 3:

There are two versions of the binary feedback scheme:

1. Destination-based

2. Source-based

In the first version, the destination examines the bits received, determines a new flow-control window, and sends this window to the source. In the second version, the destination sends all bits back to the source along with the acknowledgments. In this case, we need to reserve one bit in the headers of transport layer acknowledgment packets where the destination transport entity copies the bit received from the network layer. The source transport entity examines the stream of bits received, determines a new operating window, and uses it as long as it does not violate the window limit imposed by the the destination.

We have studied both versions. The NSP transport protocol in DNA [6] uses the source-based approach, while the ISO TP4 [9] implementation uses the destination-based approach.

In the remainder of this report, we use the word **user** to include both source and destination transport entities. Thus, when we say that the user changes its window, the change might be decided and affected by the source or destination transport entity.

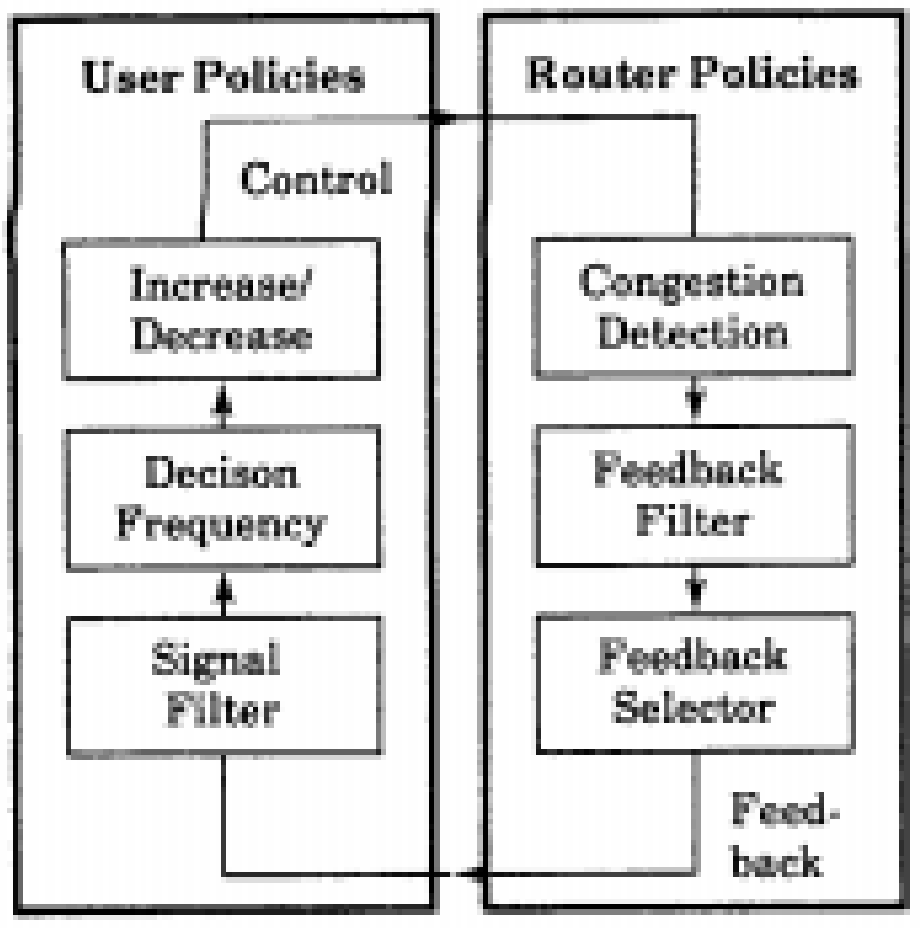

Figure 4:

The proposed congestion avoidance scheme consists of two parts: a feedback mechanism in routers, and a control mechanism for users. We call these the router policy and the user policy, respectively. Each of these mechanisms can be further subdivided into three components as shown in Figure 4. We explain these components below. For further details see [16, 22, 23].

## 5.1 Router Policies

Routers in a connectionless network environment are not informed about resource requirements of transport entities and therefore they have no prior knowledge of future traffic. They attempt to optimize their operation by monitoring the current load and by asking the users (via the bit) to increase or decrease the load. Thus, the routers have three distinct algorithms:

1. To determine the instantaneous load level

2. To estimate average load over a appropriate time interval
3. To determine the set of users who should be asked to adjust their loads

We call these three algorithms congestion detection, feedback filter, and feedback selections, respectively. The operation of these components and the alternatives considered are described next.

### 5.1.1 Congestion Detection

Before a router can feed back any information, it must determine its load level. It may be underutilized (below the knee) or overutilized (above the knee). This can be determined, based on the utilization, buffer availability, or queue lengths.

We found that the average queue length provides the best mechanism to determine if we are above or below the knee. This alternative is least sensitive to the arrival or service distributions and is independent of the memory available at the router. For both M/M/1 and D/D/1 queues the knee occurs when the average queue length is one. For other arrival patterns such as *packet trains* [14], this is approximately (though not exactly) true. The routers, therefore, monitor the queue lengths and ask users to reduce the load if the average queue length is more than one, and vice versa.

### 5.1.2 Feedback Filter

After a router has determined its load level, its feedback to users is useful if and only if the state last long enough for the users to take action based on it. A state that changes very fast may lead to confusion because by the time users become aware of it, the state no longer holds and the feedback is misleading. Therefore, we need a low-pass filter function to pass only those states that are expected to last long enough for the user action to be meaningful.

This consideration rules out the use of *instantaneous* queue lengths to be used in congestion detection. An instantaneous queue length of 100 may not be a problem for a very fast router but may be a problem for a slow router. We need to average the queue lengths over a long interval. The key question is how long an interval is long enough.

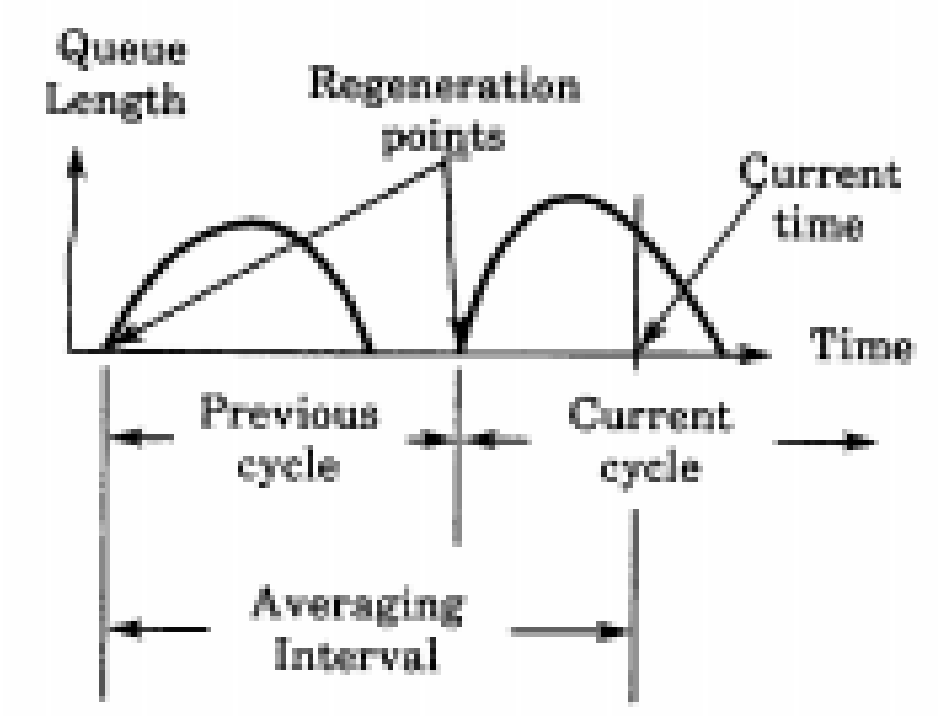

Figure 5:

We recommend averaging since the beginning of the previous regeneration cycle. A regeneration cycle is defined as the interval consisting of a *busy period* and an *idle period*, as shown in Figure 5. The beginning of the busy period is called a **regeneration point**. The word *regeneration* signifies the birth of a *new* system, since the queuing system's behavior after the regeneration point does not depend upon that before it. The average queue length is given by the area under the curve divided by the time since the last but one regeneration point. Note that the averaging includes a part of the current, though incomplete, cycle. This is shown in Figure 5.

### 5.1.3 Feedback Selection

The two components of router policy discussed so far (congestion detection and feedback filter) ensure that the router operates efficiently, that is, around the knee. They both work based upon the total load on the router, to decide if the total load is above the knee or below the knee. The total number of users or the fact that only a few of the users might be causing the overload is not considered in those components. Fairness considerations demand that only those users who are sending more than their *fair share* should be asked to reduce their load, and others should be asked to increase if possible. This is done by the feedback selection, an important component of our scheme. Without the selection, the system may stabilize at (operate around) an operating point that is efficient but not fair. For example, two users sharing the same path may keep operating at widely different throughputs.

The feedback selection works by keeping a count of the number of packets sent by different users since the beginning of the queue averaging interval. This is equivalent to monitoring their throughputs. Based on the total throughput, a fair share is determined and users sending more than the fair share are asked to reduce their load while the users sending less than the fair share are asked to increase their load. Of course, if the router is operating below the knee, each one is encouraged to increase regardless of their current load. The fair share is estimated by assuming the capacity to be at 90% of the total throughput since the beginning of the last regeneration cycle.

The feedback selection as proposed here attempts to achieve fairness among different network layer service access point (**NSAP**) pairs because the packet counts used in the algorithm correspond to these pairs.

This completes the discussion on the router policies. We now turn to the user policies.

## 5.2 User Policies

Each user receives a stream of congestion avoidance bits, called *signals*, from the network. These signals are not all identical (or else we would not need all of them). Some signals ask the user to reduce the load, while others ask it to increase the load. The user policy should be designed to compress this stream into a single increase/decrease decision at suitable intervals. The key questions that the user policy helps answer are:

1. How can all signals received be combined?
2. How often should the window be changed?
3. How much should the change be?

We call these three algorithms *signal filter*, *decision frequency*, and *increase/decrease algorithm*, respectively.

### 5.2.1 Signal Filter

The problem solved by this component is to examine the stream of the last $n$ bits, for instance, and to decide whether the user should increase or decrease its load (window). Mathematically,

$$d = f(b_1, b_2, b_3, \ldots, b_n)$$

Here, $d$ is the binary decision ($0 \Rightarrow$ increase, $1 \Rightarrow$ decrease) and $b_i$ is the the $i^{th}$ bit with $b_n$ being the most recently received bit. The function $f$ is the signal filter function. The function that we finally chose requires counting the number of 1s and 0s in the stream of the last n bits. Let

$$s_1 = \text{number of ones in the stream} = \sum b_i$$

$$s_0 = \text{number of zeros in the stream} = n - s_1$$

Then, if $s_1 > pn$ then $d = 1$ else $d = 0$. Here, $p$ is a parameter called **cutoff probability**. We found that for exponentially distributed service times, the optimal choice was $p = 0.5$, as expected. For deterministic service times, however, we found that the choice of $p$ does not matter. This is because in deterministic cases, the router filtering results in the user consistently receiving either all 1s if the load at the bottleneck is above the knee or all 0s if the load is below the knee. Based on this observation, we recommend using a cutoff probability of 50%.

In summary, the signal filtering simply consists of comparing the counts of 1s and 0s received in the bit stream and deciding to go up or down as indicated by the majority of the bits.

### 5.2.2 Decision Frequency

The decision frequency component of the user policy consists of deciding how often to change the window. Changing it too often leads to unnecessary oscillations, whereas changing it infrequently leads to a system that takes too long to adapt.

System control theory tells us that the optimal control frequency depends upon the feedback delay – the time between applying a control (change window) and getting feedback (bits) from the network corresponding to this control.

In computer networks, it takes one round-trip delay to affect the control, that is, for the new window to take effect and another round-trip delay to get the resulting change fed back from the network to the users. This leads us to the recommendation that windows should be adjusted once every two round-trip delays (two window turns) and that only the feedback signals received in the past cycle should be used in window adjustment, as shown in Figure 6.

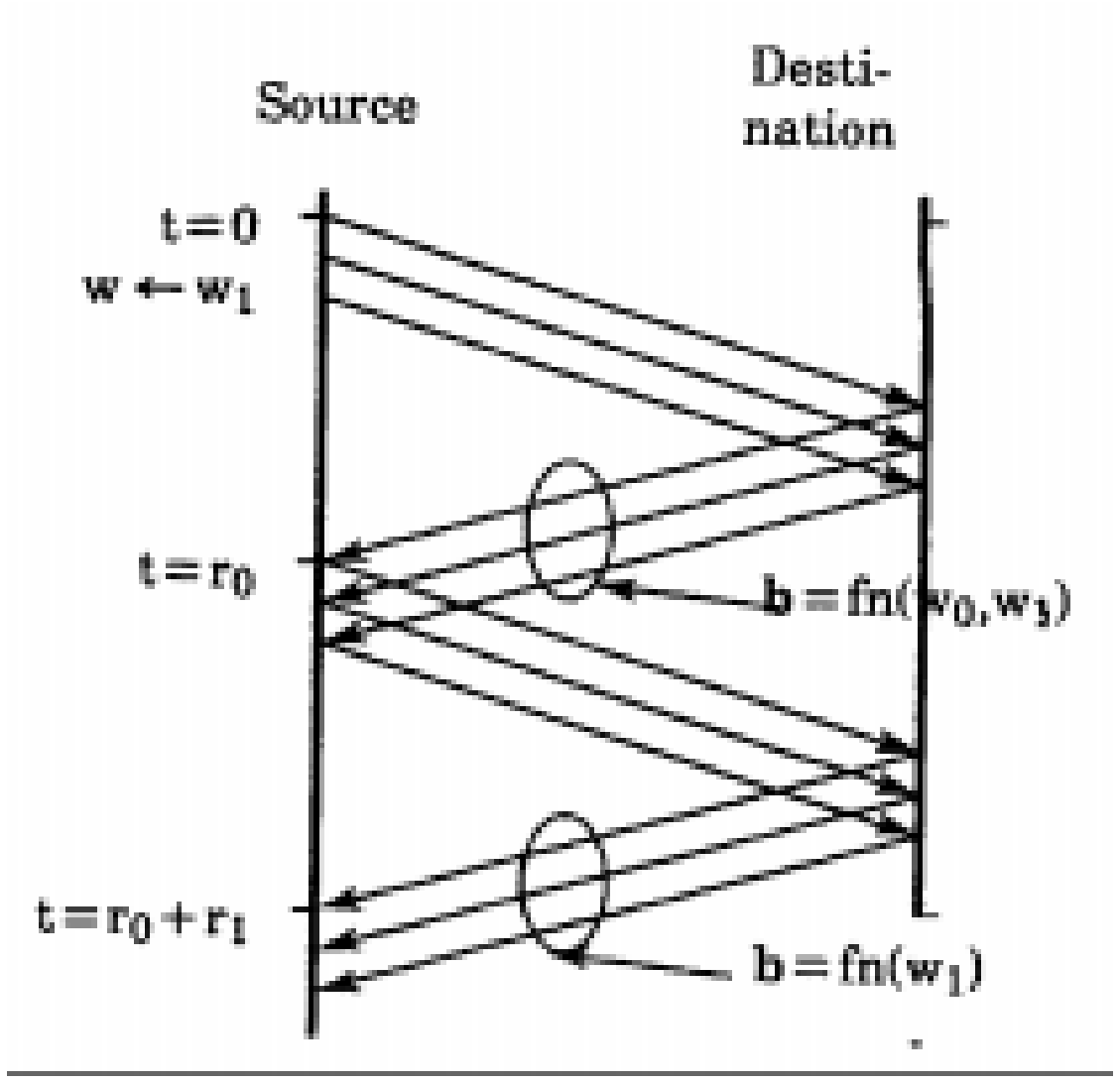

Figure 6:

#### 5.2.3 Increase/Decrease Algorithms

The purpose of the increase/decrease algorithm is to determine the amount by which the window should be changed once a decision has been made to adjust it.

In the most general case, the increase (or decrease) amount would be a function of the complete past history of controls (windows) and feedbacks (bits). In the simplest case, the increase/decrease amount would be a function only of the window used in the last cycle and the resulting feedback. Actually, there is little performance difference expected between the simplest and the most general control approach, provided that the simple scheme makes full use of the new information available since the last activation of the component. We therefore chose the simple approach. We have already partitioned the problem so that the signal filter looks at the feedback signals and decides whether to increase or decrease. The increase/decrease algorithm, therefore, needs to look at the window in the last cycle and decide what the new window should be. We limited our search among alternatives to the first order linear functions for both increase and decrease:

Increase: $w_{new} = aw_{old} + b$

Decrease: $w_{new} = cw_{old} - d$

Here, $w_{old}$ is the window in the last cycle and $w_{new}$ is the window to be used in the next cycle; $a$, $b$, $c$, and $d$ are non-negative parameters. There are four special cases of the increase/decrease algorithms:

**a** Multiplicative Increase, Additive Decrease ($b = 0$, $c = 1$)

**b** Multiplicative Increase, Multiplicative Decrease ($b = 0$, $d = 0$)

**c** Additive Increase, Additive Decrease ($a = 1$, $c = 1$)

**d** Additive Increase, Multiplicative Decrease ($a = 1$, $d = 0$)

The choices of the alternatives and parameter values are governed by the following goals:

1. Efficiency: The system bottleneck(s) should be operating at the knee.

2. Fairness: The users sharing a common bottleneck should get the same throughput.

3. Minimum Convergence Time: Starting from any state, the network should reach the optimal (efficient as well as fair) state as soon as possible.

4. Minimum Oscillation Size: Once at the optimal state, the user windows oscillate continuously below and above this state. The parameters should be chosen such that the oscillation size is minimal.

These considerations lead us to the following recommendation for increase/decrease algorithms [16, 4]:

$$\text{Additive Increase: } w_{new} = w_{new} + 1$$

$$\text{Multiplicative Decrease: } w_{new} = 0.875 w_{old}$$

If the network is operating below the knee, all users go up equally, but, if the network is congested, the multiplicative decrease makes users with higher windows go down more than those with lower windows, making the allocation more fair. Note that $0.875 = 1 - 2^{-3}$. Thus, the multiplication can be performed without floating point hardware, by simple logical shift instructions.

The computations should be rounded to the nearest integer. Truncation, instead of rounding, results in lower fairness.

This completes our discussion of the proposed binary feedback scheme. The key router and user policy algorithms are summarized in the appendix.

# 6 FEATURES OF THE SCHEME

The design of the binary feedback scheme was based on a number of goals that we had determined beforehand. Below, we show how the binary feedback scheme meets these goals.

1. No control during normal operation: The scheme does not cause any extra overhead during normal (underloaded) conditions.

2. No new packets during overload: The scheme does not require generation of new messages (e.g., source quench) during overload conditions.

3. Distributed control: The scheme is distributed and works without any central observer.

4. Dynamism: Network configurations and traffic vary continuously. Nodes and links come up and down and the load placed on the network by users varies widely. The optimal operating point is therefore a continuously moving target. The proposed scheme dynamically adjusts its operation to the current optimal point. The users continuously monitor the network by changing the load slightly below and slightly above the optimal point and verify the current state by observing the feedback.

5. Minimum oscillation: The increase amount of 1 and decrease factor of 0.875 have been chosen to minimize the amplitude of oscillations in the window sizes.

6. Convergence: If the network configuration and workload remain stable, the scheme brings the network to a stable operating point.

7. Robustness: The scheme works under a noisy (random) environment. We have tested it for widely varying service-time distributions.

8. Low parameter sensitivity: While comparing various alternatives, we studied their sensitivity with respect to parameter values. If the performance of an alternative was found to be very sensitive to the setting of a parameter value, the alternative was discarded.

9. Information entropy: Information entropy relates to the use of feedback information. We want to get the maximum information across with the minimum amount of feedback. Given one bit of feedback, information theory tells us that the maximum information would be communicated if the bit was set 50% of the time.

10. Dimensionless parameters: A parameter that has dimensions (length, mass, time) is generally a function of network speed or configuration. A dimensionless parameter has wider applicability. Thus, for example, in choosing the increase algorithm we preferred increasing the window by an absolute amount of $k$ packets rather than by a rate of $t$ packets/second. The optimal value of the latter depends upon the link bandwidth. All parameters of the proposed scheme are dimensionless, making it applicable to networks with widely varying bandwidths.

11. Configuration independence: We have tested the scheme for many different configurations of widely varying lengths and speeds including those with and without satellite links.

Most of the discussion in this and associated reports centers around window-based flow-control mechanisms. However, we must point out that this is not a requirement. The congestion avoidance algorithms and concepts can be easily modified for other forms of flow control such as rate-based flow control, in which the sources must send at a rate lower than a maximum rate (in packets/second or bytes/second) specified by the destination. In this case, the users would adjust rates based on the signals received from the network.

# 7 COMPARISON WITH SIMILAR SCHEMES

It must be pointed out that the binary feedback scheme proposed here is different from most other schemes in that it is the first attempt to achieve congestion avoidance rather than congestion control. Similar congestion control schemes exist in literature. For example, the congestion control scheme used in SNA [1] also uses bits in the network layer headers to feed back congestion information from the network to the source. It uses two bits called the *change window indicator* (CWI) and the *reset window indicator* (RWI). The first bit indicates moderate congestion, while the second one indicates severe congestion. The CWI bit is set by a router when it finds that more than a percentage, such as 75%, of its buffers have been used. After all buffers are used up, the router starts setting RWI bits in the packets going in the reverse direction. On receipt of a CWI, the source decreases the window by 1. On the receipt of a RWI, the source resets the window to $h$, where $h$ is the number of hops. If both bits are clear, the window is increased by one until a maximum of $3h$ is reached.

The key difference between SNA's scheme (and all prior work in congestion control) and our binary feedback scheme is the definition of the goal. SNA's goal is to ensure that packets find buffers when they arrive at the routers. Our scheme, on the other hand, is not so much concerned with the buffers. Rather it tries to maximize the throughput while also minimizing the delay. The routers start setting the bits as soon as the average queue length is more than one. The number of buffers available at the router has no effect on our scheme.

The key test to decide whether a particular scheme is a congestion control or a congestion avoidance scheme is to consider a network with all nodes having infinite memory (infinite buffers). A congestion control scheme will generally remain inactive in such a network, allowing the users to use large windows causing high response time. A congestion avoidance scheme, on the other hand, is useful even in a network with infinite memory. It tries to adjust queuing in the network so that a high throughput and a low response time is achieved.

# 8 PERFORMANCE

The binary feedback scheme was designed using a simulation model that allowed us to compare various alternatives and study them in detail. Most of the choices discussed earlier in this report have been justified using analytical arguments. However, we have verified all arguments using simulation as well. The model allows us to simulate any number of users going through various paths in the network. It is an extension of the model described in [12]. The model simulates portions of network and transport layers. The transport layer is modeled in detail. The routers are modeled as single server queues. The model's key limitation currently is that the acknowledgments returning from a destination to a source are not explicitly simulated. Instead, the source is informed of the packet delivery as soon as the packet is accepted by the destination.

In this section, we present a few cases to illustrate the performance of the binary feedback scheme. Other simulation results including those for random service times and highly congested networks are given in [22, 23].

### 8.1 Case I (Single User)

This case consists of a single user using a path consisting of four routers, as shown in Figure 7a.

The service times at the routers are 2, 5, 3, and 4 units of time, respectively. In our simulation the user's speed is one packet per unit of time. In other words all times are expressed as multiples of time required to send one packet. The third router is a satellite link having a fixed delay of 62.5 units of time. The second router is the bottleneck, and its power as a function of the window size is shown in Figure 7b. This graph is obtained by running the simulation without the binary feedback scheme at a fixed window and observing the user throughput and response times. It is seen from this figure that the knee occurs at a window of 15.5.

Figure 7c shows a plot of a user's window with the binary feedback scheme. The time is shown along the horizontal axis. Notice that the user starts with a window of 1 and sends packet 1 and 2; both packets traverse the subnet with the congestion bit clear. The user, therefore, increases the window to 2. Packets 3 and 4 are sent. After their acknowledgment, packets 5 and 6 are sent. The congestion bits in packets 5 and 6 are examined. They are clear and so the window is increased to 3. This continues until the window is 16. At this point, the bottleneck starts operating above the knee and starts setting congestion bits in packets. The user, upon receiving these packets, reduces the window to 16(0.875) or 13. The cycle then repeats and the window keeps oscillating between 13 and 16.

This case illustrates the fact that the network operates efficiently.

### 8.2 Case II (Two Users)

To illustrate the fairness aspects of the scheme, consider the same configuration as in Case 1, except that we have now added another user that enters the subnet at router 1 and exits after router 2 (see Figure 8a). Also, the second user starts after the first one has sent 200 packets. The optimal operating point for this case can be determined by running the simulation without the congestion avoidance scheme for various combinations of window sizes for the two users and finding the window values that are efficient and fair. The router 2 is the bottleneck. At the knee its throughput is 1/5 packets per unit time, divided equally between the two users. The optimal window in this case is $w_1 = 10$ for user 1 and $w_2 = 5$ for user 2. The plots of the two user's windows are shown in Figure 8b. Notice that with only one user the system stabilizes at the window of 15 and keeps oscillating around it until the second user joins the network. At this point, the first user receives decrease signals from the bottleneck router while the second user receives increase signals. The windows eventually stabilize when they reach their optimal values. Figure 8c shows a plot of throughputs of the two users. The throughput of the first user drops while that of the second increases until they both share the bottleneck approximately equally.

Another feature of the scheme, which can be seen from this case, is that the scheme adapts as the number of users in the network changes. The users need not start at the same point (window of 1) to reach the fair operating point.

## 9 SUMMARY

The key contributions of our congestion avoidance research are the following:

1. We have introduced the new term *congestion avoidance*. It has been distinguished from other similar terms of flow control and congestion control. It has been shown that the preventive mechanism, congestion avoidance, helps the network use its resources in an optimal manner.

2. We defined the concept of *global optimality* in a distributed system with multiple resources and multiple users. The optimality is defined by efficiency and fairness. Both concepts have been developed so that they can be independently quantified and can apply to any number of resources and users.

3. Other researchers attempting to define global optimality have had difficulty extending the concept of power to distributed resources. By defin-ing efficiency for each resource and relating fairness to users, we have been able to separate the two concepts.

4. We have developed a simple scheme that allows a network to reach the optimal operating point automatically. This scheme makes use of a single bit in the network layer header. This bit is shared by all resources.

5. We divided the problem of congestion avoidance into six components which can be studied separately. This allowed us to compare a number of alternatives for each component and select the best.

6. We have simulated the binary feedback scheme and tested its performance in many different configurations and conditions. The scheme has been found to operate optimally in all cases tested.

# 10 ACKNOWLEDGMENTS

Many architects and implementers of Digital's networking architecture participated in a series of meetings over the last three years in which the ideas presented here were discussed and improved. Almost all members of the architecture group contributed to the project in one way or another. In particular, we would like to thank Tony Lauck and Linda Wright for encouraging us to work in this area. Radia Perlman, Art Harvey, Kevin Miles, and Mike Shand are the responsible architects whose willingness to incorporate our ideas provided further encouragement. We would also like to thank Bill Hawe, Dave Oran, and John Harper for feedback and interest. The idea of proportional decrease was first proposed by George Verghese and Tony Lauck. The concept of maximal fairness was developed by Bob Thomas and Cuneyt Ozveren.

**Appendix A: Algorithms**

The SIMULA procedures used in the simulation model are included here to clearly explain various algorithms used in the scheme. The data structures used by the queue servers in the routers are presented followed by five procedures which are used as follows:

1. Arrival: This procedure is executed on each packet arrival. It computes the area under the queue length curve. Also, at the begining of a new cycle, the tables are initialized. SIMULA variable 'time' gives the currently simulated time.

2. Departure: This procedure is executed on the packet departure. It decreases the queue size and updates the value of area under the queue length curve. A hash function is used to find the table entry where counts for packets sent by this user are kept.

3. Fair_Share: This procedure is used to decide the maximum number of packets any user should be

allowed to send. The routers set the congestion avoidance bits in packets belonging to users sending more than this amount.

4. Increase: This procedure is used by a transport entity to increase its window if less than 50% of the bits received are set.

5. Decrease: This procedure is used by a transport entity to decrease its window if more than or equal to 50% of the bits are set. The SIMULA function Entier(x) returns the highest integer less than or equal to x.

The first two procedures are parts of the *feedback filter* algorithm discussed earlier under router policies. The third procedure constitutes the *feedback selector* algorithm. The last two procedures make up the *increase/decrease* algorithms of the user policies.

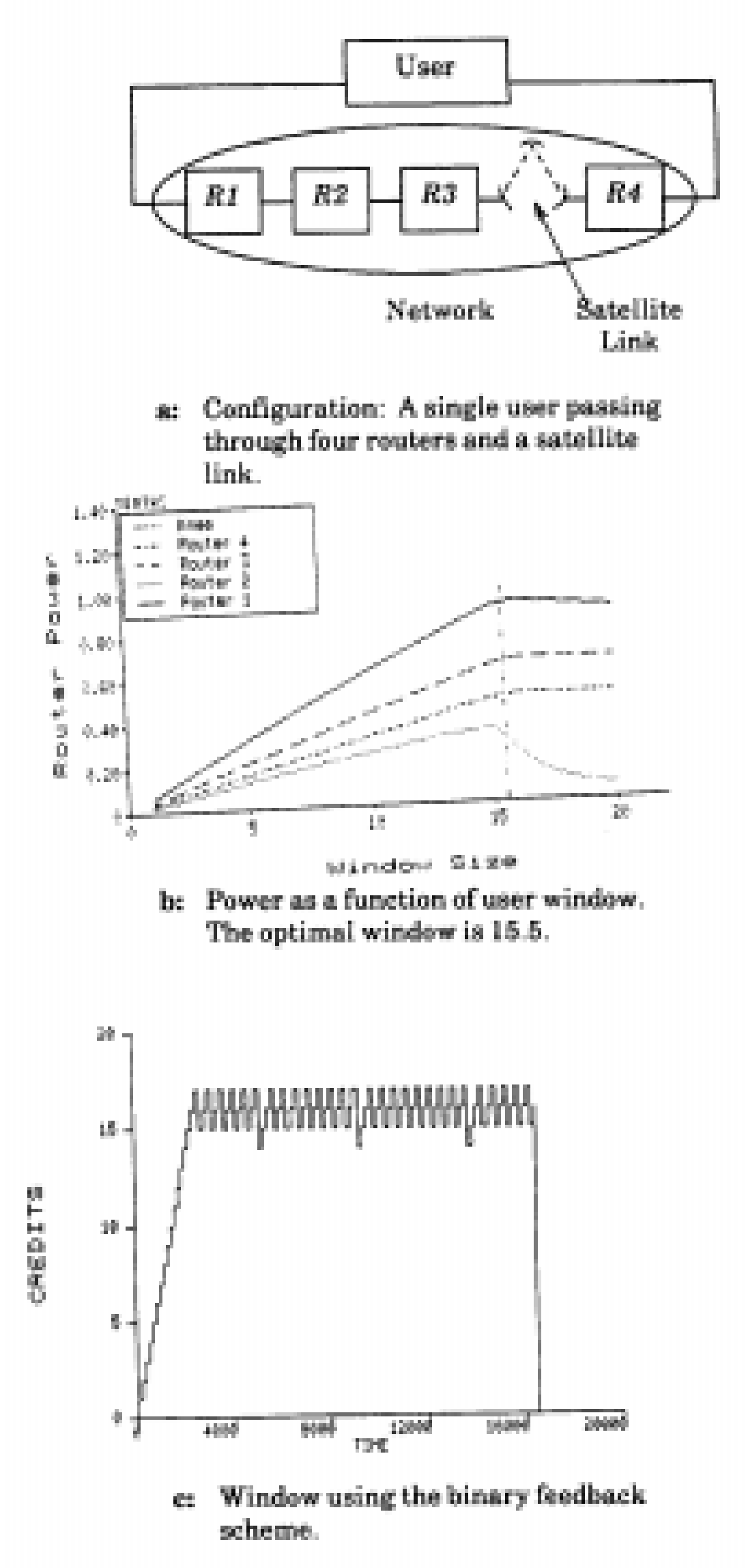

a: Configuration: A single user passing through four routers and a satellite link.

b: Power as a function of user window. The optimal window is 15.5.

c: Window using the binary feedback scheme.

Figure 7:

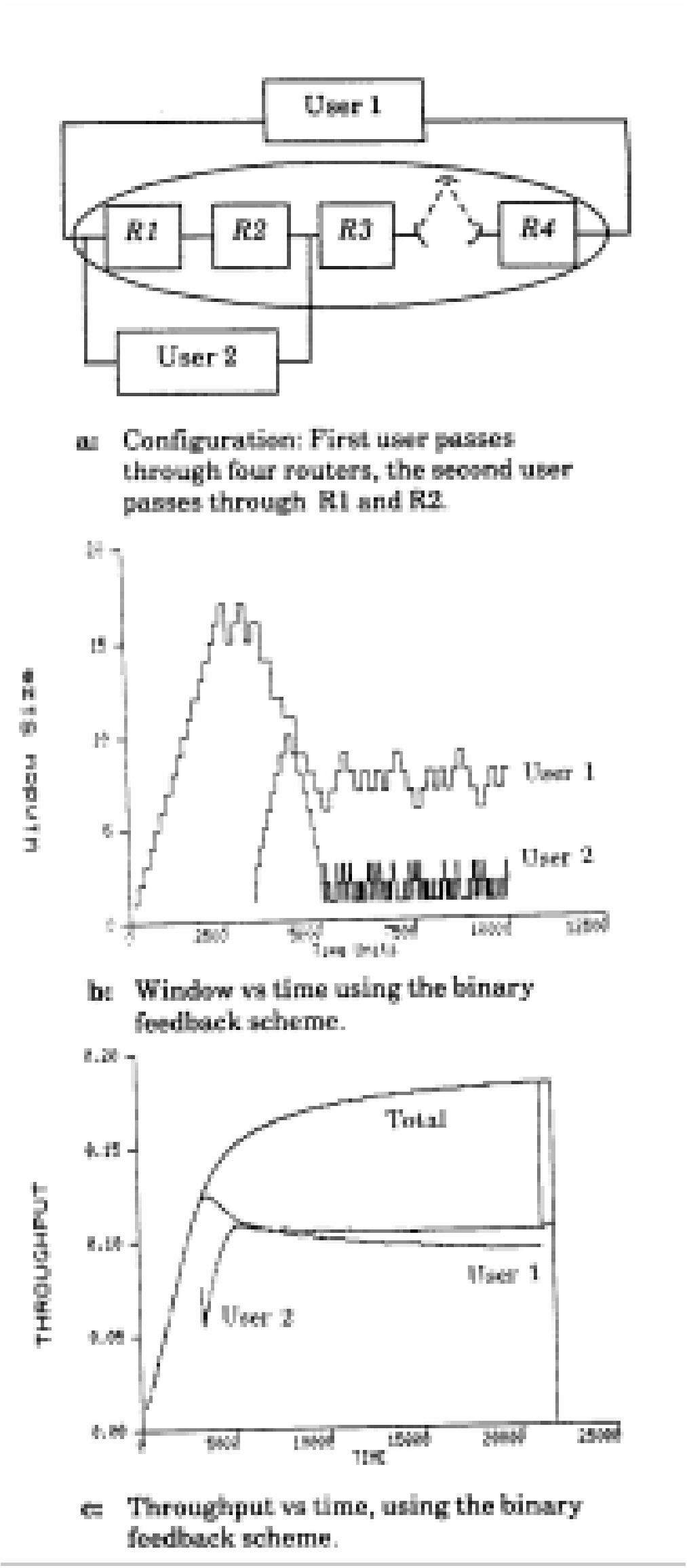

a: Configuration: First user passes through four routers, the second user passes through R1 and R2.

b: Window vs time using the binary feedback scheme.

c: Throughput vs time, using the binary feedback scheme.

Figure 8:

```
!The following data structure is maintained by each queue server or router.
 The size of the table 'dim_tables' to be used is left to the implementors;
   REAL ARRAY packets_sent[0:dim_tables];!Table for keeping packet counts;
                                    !0th location is used for total count;
   REAL ARRAY prev_packets_sent[0:dim_tables]; !Counts for previous cycle;
   REAL avg_q_length;               !Average queue length at this server;
   REAL area;                       !Area under Q length vs time curve;
   REAL prev_area;                  !Area in the previous cycle;
   INTEGER q_length;                !Queue length includes one in service;
   REAL q_change_time;              !Last time the queue changed;
   REAL prev_cycle_begin_time;      !Time at which previous cycle began;
   REAL cycle_begin_time;           !Time at which this cycle began;

   PROCEDURE arrival;               !To be executed on packet arrival;
   BEGIN
       INTEGER i;                   !Temporary index variable;
       area:=area+q_length*(time-q_change_time);!Compute Area under the curve;
       q_length:=q_length+1;        !Increment number in the queue;
       q_change_time:=time;         !Time of change in Queue length;
       IF(q_length=1) THEN          !Begining of a new cycle;
       BEGIN                        !End the previous cycle;
           prev_cycle_begin_time:=cycle_begin_time;
           cycle_begin_time:=time;
           prev_area:=area;
           area:=0;
           FOR i:=0 STEP 1 UNTIL dim_tables DO
           BEGIN
               prev_packets_sent[i]:=packets_sent[i];!Remember all counts;
               packets_sent[i]:=0;  !Clear packet counts;
           END;                     !of FOR;
       END;                         !of IF(Q_length=1);
   END of arrival;

   PROCEDURE departure;             !To be executed on packet departure;
   BEGIN
       BOOLEAN bit;                 !The congestion avoidance bit in the packet;
       INTEGER user;                !Index in the packet count table;
       area:=area+q_length*(time-q_change_time);!Compute area under the curve;
       q_length:=q_length-1;        !Decrement the number in the queue;
       q_change_time:=time;         !Rememeber time of queue length change;
       avg_q_length:=(area+prev_area)/(time-prev_cycle_begin_time);!Compute avg Q length;
       user:=hash(source_address,dest_address,dim_tables);!Find index into the table;
       packets_sent[user]:=packets_sent[user]+1;!Increment the count;
       packets_sent[0]:=packets_sent[0]+1;!Increment total count also;
       IF(avg_q_length>2)           !Are we heavily congested?;
       THEN bit:=TRUE               !Yes, set bit for all users;
```

```
    ELSE IF (avg_q_length<1) THEN   !No,  do nothing if we are underloaded;
    ELSE IF packets_sent[user]+prev_packets_sent[user]>fair_share
    THEN bit:=TRUE;                 !If the user sent too many packets, set bit;
END of departure;
```

```
REAL PROCEDURE fair_share;        !Computes the max number of packets a user can send;
BEGIN
    REAL capacity;                  !Knee capacity of the server;
    REAL old_fair_share;            !Max allocation (used previously);
    INTEGER sum_allocation;         !Total capacity allocated;
    INTEGER old_sum_allocation;     !Capacity allocated (previously);
    INTEGER i;                      !Index variable;
    REAL demand;                    !Demand of the ith user;
    INTEGER num_not_allocated;      !Number of users yet to be allocated;
    capacity := 0.9*(packets_sent[0]+prev_packets_sent[0]);
                                    !Assume capacity=90% of packets sent;
    num_not_allocated := dim_tables;!Initialize number of users to be allocated;
    sum_allocation := 0;            !Total allocation so far;
    old_sum_allocation := -1;       !Allocation in the previous iteration;
    fair_share := -1;               !Users below this allocation are good;
    WHILE (sum_allocation>old_sum_allocation) DO
    BEGIN                           !Beginning of a new iteration;
        old_fair_share := fair_share;
        old_sum_allocation := sum_allocation;
        fair_share := (capacity-sum_allocation)/num_not_allocated;!New estimate;
        FOR i := 1 STEP 1 UNTIL dim_tables DO
        BEGIN
            demand:=packets_sent[i]+prev_packets_sent[i];!Demand in the last two cycles;
            IF(demand<=fair_share AND demand>old_fair_share)
            THEN BEGIN
                num_not_allocated := num_not_allocated-1;!One more user satisfied;
                sum_allocation := sum_allocation+demand;
            END;                    !of IF;
        END;                        !of FOR;
    END;                            !of WHILE;
END of fair_share;

PROCEDURE increase(w,w_max,w_used); !Used to increase the window;
NAME w,w_used;                      !These parameters are called by name;
REAL w;                             !Computed window (real valued);
INTEGER w_max;                      !Max window allowed by the destination;
INTEGER w_used;                     !Window valued used (integer valued);
BEGIN
    w:=w+1;                         !Go up by 1;
    IF(w>(w_used+1)) THEN w:=w_used+1;!No more than 1 above the last used;
    IF w>w_max THEN w:=w_max;       !Also, never beyond the destination limit;
    w_used:=Entier(w+0.5)           !Round-off;
END of increase;

PROCEDURE decrease(w,w_used);       !Used to decrease the window;
NAME w,w_used;                      !These parameters are called by name;
```

```
REAL w;                             !Computed window (real valued);
INTEGER w_used;                     !Window value used (integer valued);
BEGIN
    w:=0.875*w;                     !Multiplicative decrease;
    IF(w<1)THEN w:=1;               !Do not reduce below one;
    w_used:=Entier(w+0.5);          !Round-off;
END of decrease;
```